# A Systematic Mapping Study on Open Source Agriculture Technology Research

**Kevin Lumbard**

Creighton University, USA

kevinlumbard@creighton.edu

**Vinod Kumar Ahuja**

Florida Gulf Coast University, USA

vahuja@fgcu.edu

**Matt Cantu Snell**

University of Nebraska at Omaha, USA

matt@cantus.email

**Abstract:**

Agriculture contributes trillions of dollars to the US economy each year. Digital technologies are disruptive forces in agriculture. The open source movement is beginning to emerge in agriculture technology and has dramatic implications for the future of farming and agriculture digital technologies. The convergence of open source and agriculture digital technology is observable in scientific research, but the implications of open source ideals related to agriculture technology have yet to be explored. This study explores open agriculture digital technology through a systematic mapping of available open agriculture digital technology research. The study contributes to Information Systems research by illuminating current trends and future research opportunities.

# 1   Introduction

Agriculture contributes trillions of dollars to the US economy each year (Leply, 2019). Increasingly, modern farms utilize advanced digital technology to increase efficiency and productivity. Digital technology in agriculture encompasses a wide variety of hardware and software systems (Weltzien, 2016). The use of digital technology in agriculture, often referred to as smart farming or precision farming, aims to increase production and sustainability of resources through the adoption of digital technology (Yost et al., 2019). Agriculture digital technology is a collection of frequently connected technologies including sensor systems, Information Systems, enhanced machinery, decision support, and supply chain systems (Gebbers & Adamchuk, 2010).

Digital technologies are disruptive forces in agriculture. They result in the creation of new business models, organizational changes, new forms of innovation, and the requirement of new skill sets (Shepherd et al., 2018). Technological changes to the agriculture industry are expected to cause shifts in the roles and power relations of agriculture stakeholders (Wolfert et al., 2017). The adoption of digital technology in agriculture is a response to higher profit and yield potentials; however, the way these technological changes affect individual farmers, agriculture policy, and the agriculture industry is unclear (Schieffer & Dillon, 2015; Schimmelpfennig, 2016). Despite uncertainties, precision farming is an essential trend for sustainable agriculture as farmers deal with growing demand and climate change (Arslan et al., 2015; Walter et al., 2017).

Agriculture digital technology has historically been proprietary, and they are controlled by large corporations such as Cargill, John Deere, DowDuPont, and Bayer. However, the open source movement is beginning to emerge in agriculture technology and has dramatic implications for the future of farming and agriculture digital technologies. Not just a licensing designation, open source is a method of collaboration and decentralized production that began with software and expanded to many fields, including hardware, systems, and standards. It enables open collaboration and the sharing of technologies through permissive licensing (Gamalielsson & Lundell, 2017). Open agriculture refers to the application of the open source ideas to agriculture technology. Open agriculture is observable in practice through agriculture stakeholder adoption and contributions to open source projects such as the Linux Kernel [1] (operating system), Hyperledger[2] (blockchain), and Dronecode[3] (autonomous vehicle software). One open source project that focuses on the agriculture industry is Farm OS, which provides an open source suite of tools to help farmers manage their enterprises. Another example is Farm Hack[4], a practitioner website which promotes the development of open source software and hardware technology, including open source tractors and other open source farm equipment.

The convergence of open source and agriculture digital technology is observable in scientific research, but the implications of open source ideals related to agriculture technology have yet to be explored. Both open source (Fitzgerald, 2006; Germonprez et al., 2017; von Krogh et al., 2012) and digital agriculture technology (Lokuge et al., 2016; Nugawela & Sedera, 2020; Power & Hadidi, 2019) are highly relevant to Information Systems (IS) research. Despite this relevance, Information Systems research on agriculture digital technology has been limited, and Information Systems researchers have yet to fully explore this emerging topic in the context of open source.

This study explores open agriculture digital technology through a systematic mapping of available open agriculture digital technology research. This mapping study has the following objectives:

> **Objective 1: To** develop an understanding of how open technologies and processes are utilized in agriculture digital technology based on existing research
>
> **Objective 2:** To get an overview of the current state of research on open agriculture digital technology applications
>
> **Objective 3:** To identify promising directions for future Information Systems research

---

1 https://github.com/torvalds/linux
2 https://www.hyperledger.org/
3 https://www.dronecode.org/
4 https://farmhack.org/tools

A systematic mapping study aims to get a rigorous and detailed overview of a research topic, identify research gaps, and provide insight into future research directions (Li et al., 2015). Our systematic mapping study offers an empirical examination of open agriculture digital technology research using rigorous methods to ensure validity and reliability (Li et al., 2015). While the study endeavors to be comprehensive through an extensive and multidisciplinary database search, it is not intended to be definitive. Further, this systematic mapping study has not undertaken a critical review of existing research (Grant & Booth, 2009), which evaluates the quality of papers included in the research.

The paper is structured as follows: Section two describes the research questions of the systematic mapping study. Section three describes the methods employed and presents the results mapped to the objectives and research questions. Section four discusses the results and implications for open agricultural practice and Information Systems research discipline. Section five explains the limitations and threats to validity. Section six presents the conclusion and call to action for Information Systems researchers.

## 2    Research Questions

**Objective 1:** To develop an understanding of how open source technologies and processes are utilized in agriculture technology based on existing research

The RQs in this objective explore the different types of digital technologies used in agriculture and their open source context. The use of Arduino[5] micro-controller boards is an example of open hardware technology. Blockchain technology could be associated with open software, open systems, or open standards according to the background. Thus, we propose the following two RQs:

>   **RQ1:** What types of ***open source*** are present in agriculture digital technology?

>   **RQ2:** What ***open technologies*** are being utilized in precision farming?

Open source is a licensing designation that enables open design processes. Examples of open source include software, data, hardware, systems, standards, and design processes. By answering these research questions, we can create an overview of the types of open digital technologies and open processes utilized by agricultural organizations and farmers.

**Objective 2:** To get an overview of the current state of research on open agriculture digital technology applications

The RQs in this objective will identify the specific applications of open agriculture digital technology. Answering these questions will create a list of the particular agricultural contexts within which researchers have explored precision farming. Thus, we propose the following two RQs:

>   **RQ3:** What are the different **applications** of open agriculture digital technology (production, knowledge management, specific contexts)?

>   **RQ4:** What are the different **agricultural contexts** where technology is used?

Open source technology is prevalent in technology development; however, the specific usage of open source in agricultural contexts is under-researched. By answering these research questions, we can understand the specific settings and applications that are enabled by open source licensing and practices.

**Objective 3:** To identify promising directions for future Information Systems research

RQs from objectives one and two provide insight into gaps in current research, the focal concerns of specific research disciplines, and future research directions.

## 3    Study Execution

### 3.1    Scope and selected databases

To answer our research questions, we performed a systematic mapping study (Li et al., 2015) of studies extracted from seven research databases. We did not set a start date for the search period, but the end

---

period for the search was March 2020 (when we began this study). The search extracted papers published between 2003 to 2019. We chose two databases related to computer science and Information Systems (ACM[6] and AIS[7]) as these databases are relevant to our research discipline (i.e., Information Systems) and our target contribution. Additionally, we selected four large interdisciplinary databases (Web of Science[8], IEEExplore[9], ProQuest[10], and Academic Search Complete[11]) to capture work being done in other disciplines (Webster & Watson, 2002). These database searches allowed us to capture studies from agriculture, earth sciences, and engineering research venues that would have been missed otherwise. To validate our search and to ensure that we did not miss any relevant papers, we performed a Google Scholar[12] extension search to find any studies that may have been missed in our initial search.

## 3.2    Search method and mapping results

Through discussion and exploratory searches, we identified and narrowed our search terms down to five keywords through discussion and exploratory searches. The keywords we used are open, precision, smart, farm, and agriculture. We found the words precision and smart provided accurate and extensive results. The term "Open" was included in this list of terms to find studies that explicitly mention open source concepts. From these keywords, we created a search logic and repeated it in all the selected databases. The search logic was - *(Precision or Smart) and (Agriculture or Farm) and Open*.

Our goal was to find the open source technology contexts used in agriculture digital technology. After removing duplicate studies, our initial search yielded 2,421 studies containing the combination of our search terms. The search was greedy to ensure that we did not miss relevant papers. It yielded many results, not all of which were relevant. For example, in many studies, the term Open was found to refer to open farm fields that are not related to the context of open source. Similarly, many studies were labeled as open access, but did not explicitly deal with open source technology.

Another example was the search term farm, which included studies on both wind farms and server farms. To ensure that all the papers in the systematic mapping study were relevant to open agriculture digital technology, we divided the 2,421 search results equally among the three researchers. The titles, keywords, abstracts and paper text (if necessary) were examined, and non-relevant studies were removed to ensure that the selected studies had both an explicit open and agricultural context. Non-English language studies were also removed due to ensure comprehension by the research team. The process resulted in 115 studies that uniquely presented both open source and digital agriculture components. Using the same methods described above, we then queried google scholar to validate our results and identify any missing relevant papers. We selected the top 80 results from Google Scholar. After removing duplicates and non-relevant results, we found 27 relevant studies that did not appear in our initial database searches. Together, the database search and Google Scholar extension search yielded a total of 142 relevant studies that are included in our analysis (refer to Figure 1 for a flowchart of our search and extraction process).

## 3.3    Demographics Results

This section describes the study classifications by publication type, date, and research discipline.

---

### 3.3.1 Classification by publication type

We limited our search to peer-reviewed studies that were published in conference proceedings and journals. This decision excluded content from books, workshops, and news sources allowing us to focus on the mapping results from empirical research. Interestingly, publication types appeared at almost equal frequencies. Due to the larger number of available conference venues and higher acceptance rates, we expected to see a higher number of conference papers on open agriculture digital technology in relation to journal papers. While we found more conference papers than journal papers, the difference was not as significant as we expected. In the 142 studies we extracted, 73 were conference papers and 69 were journal studies.

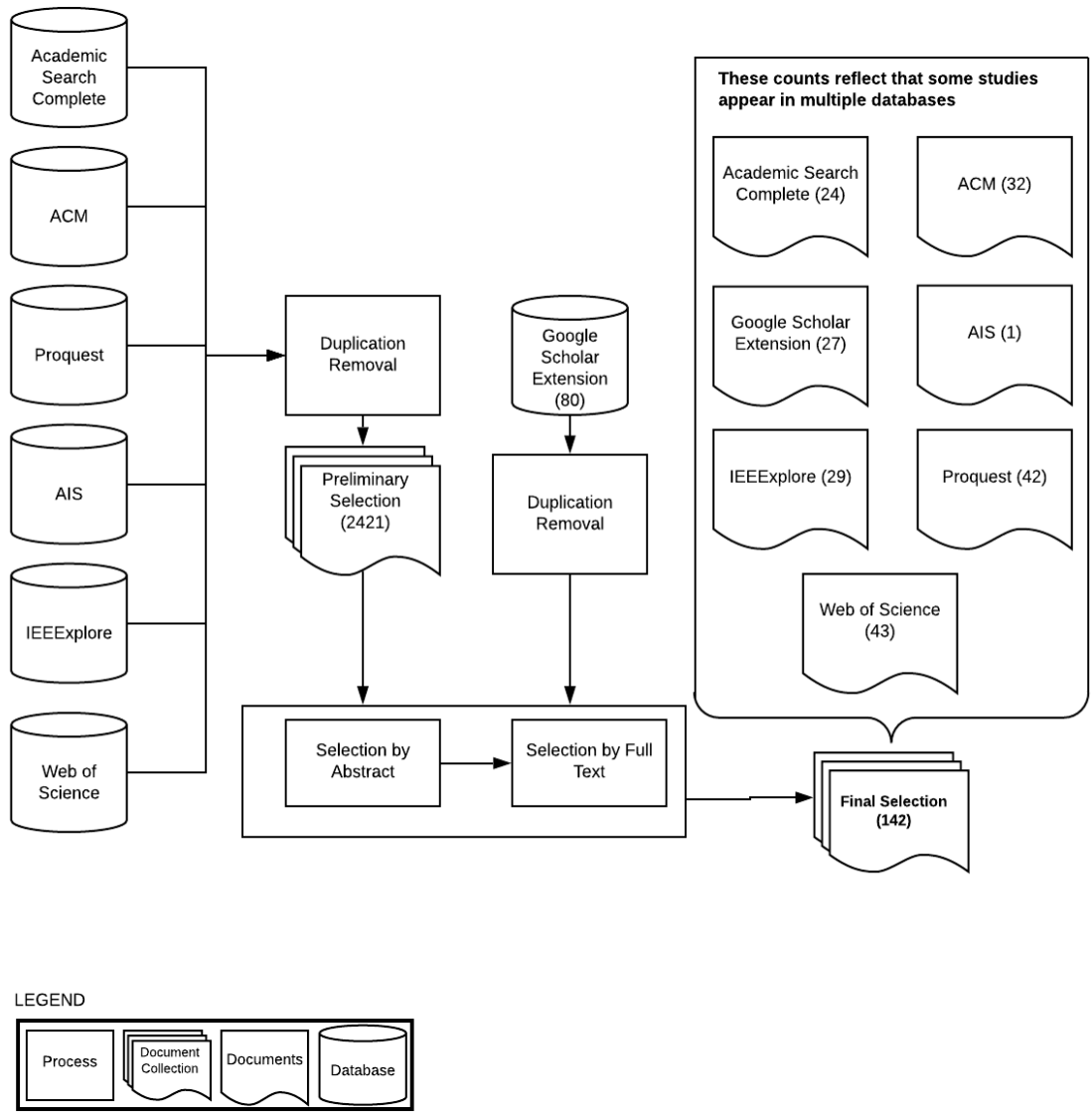

**Figure 1. Search method and selection results**

### 3.3.2    Classification by publication year

Figure 2 shows the distribution of selected studies over time. Our extraction revealed that the presence of open source in agriculture digital technology research is emergent and has started appearing only within the last decade.

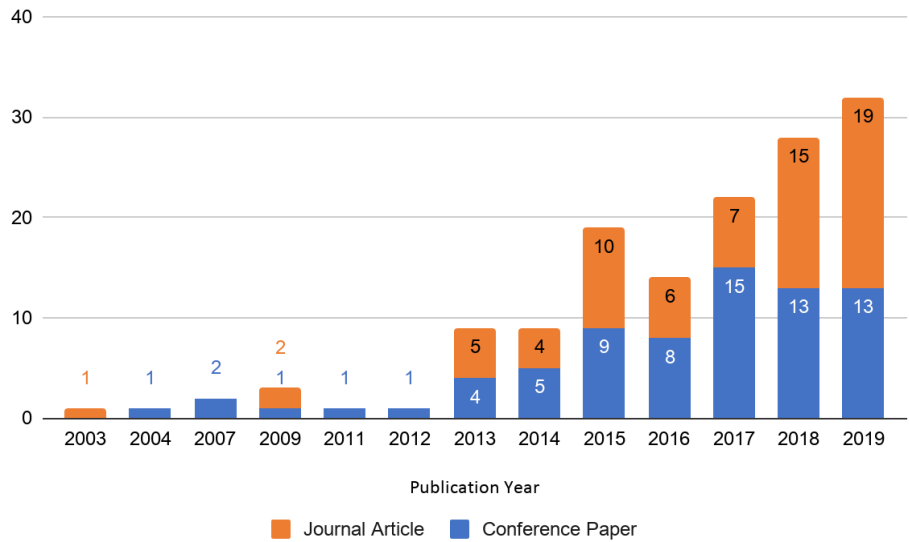

**Figure 2. Open agriculture digital technology studies by date**

### 3.3.3    Classification by research discipline

Though the presentation and results of this systematic study are targeted to Information Systems researchers, this systematic mapping study does not just explore Information Systems literature. As Webster & Watson (2002) proposed, we have included many publication venues and disciplines to help inform the development and avenues for future Information Systems research.

Each selected study from our results was categorized by research discipline. The categorization was carried out by reviewing each specific publication venue. For example, a journal focused on agriculture research was coded as "agriculture". Some journals were identified as interdisciplinary, which were categorized in all relevant disciplines. The highest number of studies (27%) appeared in the Computer Science (CS) discipline, followed by Information Systems (21%), Engineering (19%), Agriculture (17%), and Earth Sciences (16%) (Refer to Figure 3).

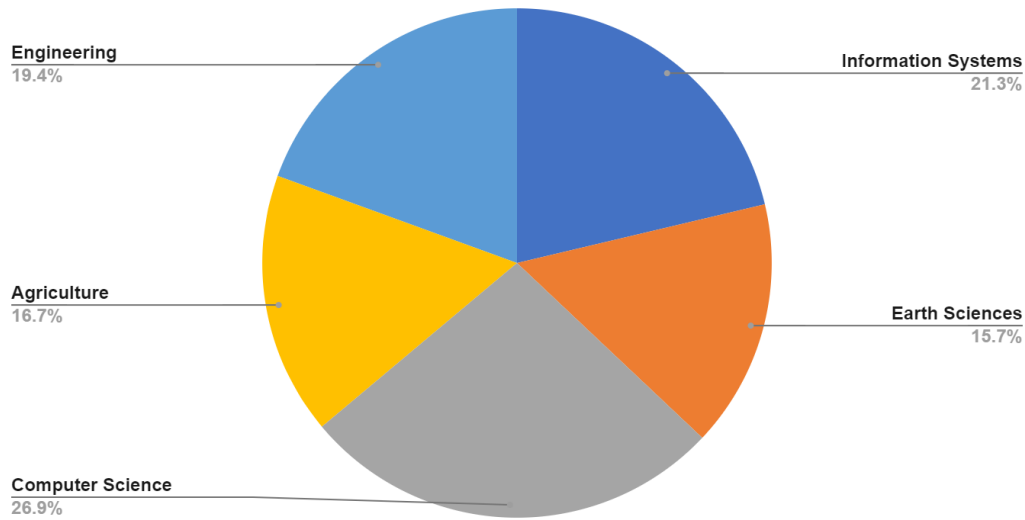

**Figure 3. Percentage breakdown of open agriculture digital technology research by discipline**

## 3.4 Qualitative Coding and Thematic Analysis

To answer our research questions, data were analyzed using qualitative coding methods (Mayring, 2004), descriptive statistics, co-occurrence matrices, and frequency analysis to find the recurring themes across the study results. For example, to answer our first question, i.e., "*what types of **open source** are present in agriculture digital technology?*," we used qualitative coding to identify open source contexts discussed in the extracted studies and performed a frequency count to identify trends and emerging applications of open source technologies.

The 142 studies were split among the three authors. Each author coded two-thirds of the studies. This disbursement allowed for overlapping assignments where each study was coded by two authors and consensus was reached through discussion. Where necessary, the third author, who was not assigned to the study, acted as a tiebreaker. The studies were coded into four overarching categories, including open source context, the type of technology used, technology applications, and specific agriculture context (refer to Figure 4).

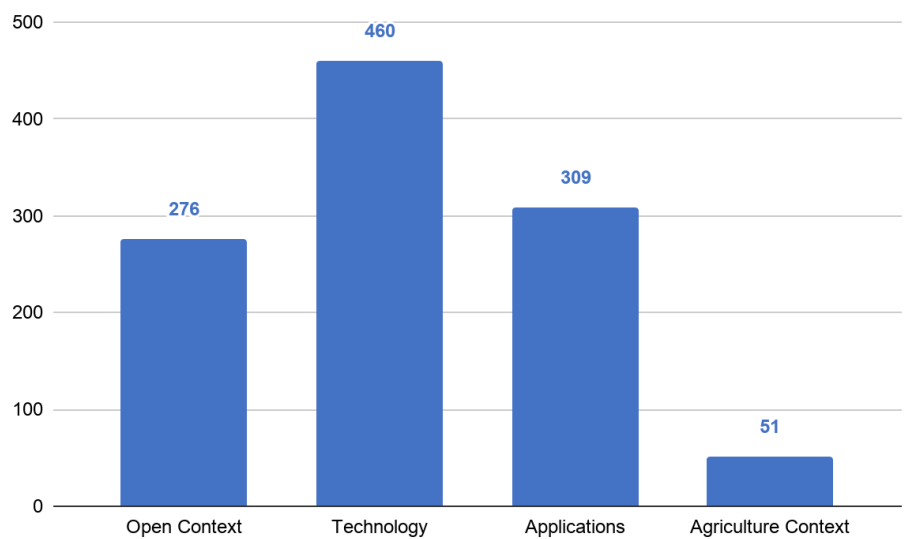

**Figure 4. Coding category distribution**

Within these categories, specific codes and themes were identified using qualitative coding content analysis methods (Mayring, 2004). In the following sections, we systematically map the observed codes to the research questions and extracted studies.

### 3.4.1 Research Question 1: Open Context

*What types of **open source** are present in agricultural digital technology research?*

Analysis of the selected studies resulted in six variations of open source context (refer to Figure 5 and Table 1). Each of the studies extracted included at least one explicit reference to open source and often included multiple contexts of open source.

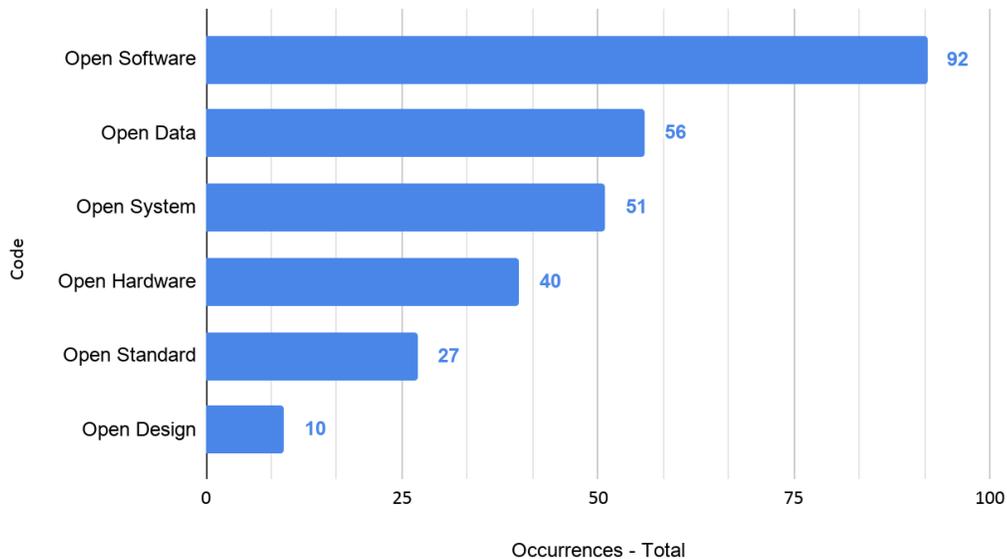

**Figure 5. Open source context codes frequency**

**Open Software** was the most observed code in our analysis and appeared in 92 studies. It was observed in studies that were using or building open source software. Open software is developed using computer mediated technology that enables the collaborative development of software artifacts by groups of distributed developers with disparate motivations (Fitzgerald, 2006; von Krogh et al., 2012). In the studies we analyzed, open software was used to run and control hardware such as generic sensors, Arduinos, and Raspberry Pi. Open software was also used for data collection and analysis in weather prediction, finding the water level in soil, and anticipating and identifying insects that can harm crops.

**Open Data** was the second most frequently occurring code and was observed in 56 studies. Open data applies the principles of open source to data, allowing communities to collaborate (Link et al., 2017). Many databases are put into the public domain allowing access and sharing of data among many farmers. An example of this is open geospatial data that is created by researchers and government entities. In this study we observed the use of open data sets and collection of open data to inform decisions based on local and external environmental factors.

**Open Systems** was coded in 51 studies. Open systems research explores the concept that Information Systems are collections of software and hardware that require the coordination of interdependencies (Lindberg et al., 2016). This code had high frequencies of occurrence with other open contexts. In agriculture an open system utilizes hardware and software to collect data and utilize it to make intelligent decisions or control processes to optimize crop yield. For Example, decision support systems, farm management systems, power management systems, and global positioning systems are developed and integrated with each other to manage farms and increase crop yield.

**Open Hardware** was identified in 40 studies. Open hardware replicates the open design processes and ideals from open software into physical technology production, specifically, the production of designs for electronic hardware and other physical objects (Powell, 2012). Open hardware can be observed as a collection of hardware pieces integrated to achieve a specific outcome and the hardware design is shared with the community so that others can benefit from it. In an agricultural context, users utilize hardware technologies such as Arduino, Zigbee, Bluetooth, Wi-Fi, Sensors, Raspberry Pi, drones, and unmanned vehicles. Many of the studies we analyzed have utilized open hardware sensors and microcontroller boards to create Internet of Things (IoT) sensor networks.

**Open Standards** were coded in 27 studies. Open standards are understood as technologies whose specifications are public and without any restriction in their access and implementation (Fomin et al., 2008). Open standards are necessary for technology applications to communicate with one another. Open standards ensure that systems are interoperable and use the same meta data specifications. In open agriculture, they are often observed in IoT systems that must communicate with multiple hardware and software technologies. They are further evident in standards for using and sharing open data such as geospatial and meteorological data.

**Open Design** was coded in 10 studies. Open design is a collaborative design process enabled by permissive licenses and computer mediated technology. Open design describes a task focused process where independent layers of work are added to open design artifacts (Howison & Crowston, 2014). Open design contributions are often motivationally separate from the overall project goals making it dynamic and responsive to changing contributors, goals, and market conditions (Germonprez et al., 2017). We found very few studies (10) that focus on open design collaboration of stakeholders in agriculture to create open artifacts. This indicates that much of the open agriculture research is focused on consumption of open technologies rather than creation. The few studies we coded with open design were focused on the creation of systems and standards for open data such as climate and geospatial information.

**Table 1. Open Context mapping**

| Code (Frequency) | Concept | Related Studies |
|---|---|---|
| Open Software (92) | Open source licensing and concepts applied to software development | [1], [2], [3], [4], [5], [7], [8], [11], [13], [14], [15], [17], [18], [19], [20], [21], [22], [24], [26], [27], [28], [29], [31], [32], [33], [34], [35], [39], [40], [42], [43], [44], [47], [49], [50], [51], [52], [53], [54], [55], [61], [63], [64], [65], [66], [68], [69], [70], [71], [72], [75], [77], [78], [79], [81], [82], [84], [85], [88], [90], [91], [92], [93], [95], [96], [100], [104], [105], [106], [108], [111], [112], [114], [115], [117], [118], [119], [120], [121], [122], [124], [126], [129], [132], [133], [134], [135], [136], [137], [140], [141], [142] |
| Open Data (56) | Data that is freely available to anyone without restrictions. This often applies to shared data sets and data created by government agencies | [3], [4], [6], [15], [17], [21], [22], [23], [25], [29], [33], [34], [36], [37], [39], [40], [43], [45], [48], [49], [51], [56], [59], [61], [63], [64], [66], [69], [70], [71], [80], [81], [82], [86], [87], [89], [90], [93], [96], [97], [98], [99], [100], [101], [102], [107], [108], [113], [115], [116], [125], [127], [130], [131], [138], [139] |
| Open System (51) | Open source licensing and concepts that encompass ecosystems or networks of open technologies | [1], [11], [12], [13], [17], [21], [22], [26], [29], [31], [34], [37], [41], [44], [45], [46], [55], [56], [57], [60], [62], [64], [66], [67], [69], [72], [75], [76], [78], [79], [80], [82], [83], [85], [89], [92], [94], [95], [96], [100], [112], [114], [117], [121], [123], [127], [131], [132], [135], [141], [142] |
| Open Hardware (40) | Open source licensing and concepts applied to hardware and physical technologies | [1], [8], [9], [16], [22], [24], [27], [30], [31], [32], [38], [42], [43], [44], [47], [57], [58], [60], [67], [73], [74], [76], [78], [82], [84], [88], [91], [94], [95], [103], [106], [107], [114], [115], [121], [122], [123], [135], [136], [137] |
| Open Standard (27) | Open source models and open standards defined through open collaboration or to enable open collaboration | [10], [21], [25], [33], [35], [39], [40], [41], [43], [62], [74], [75], [77], [80], [81], [85], [89], [96], [109], [112], [115], [118], [125], [128], [138], [139], [140] |
| Open Design (10) | Open source concepts that enable an open collaborative design process - sometimes referred to as open innovation | [12], [22], [23], [25], [36], [62], [66], [91], [110], [142] |

### 3.4.2 Research Question 2: Open Technology

*What open technologies are being utilized in precision farming?*

Analysis of the selected studies revealed 19 types of open source technology (refer to Figure 7). **IoT** and **Sensor Network** were by far the most frequently occurring technology codes. In practice, these technologies are very similar and might be descriptive of one another. Looking at the frequencies and types of codes that were identified it is easy to see that many of the technologies we observed focus on measuring environmental conditions and are often part of networked systems. These results show that most research in open agriculture digital technology is focused on IoT systems and that an assortment of different technologies are used to monitor environmental conditions in these systems.

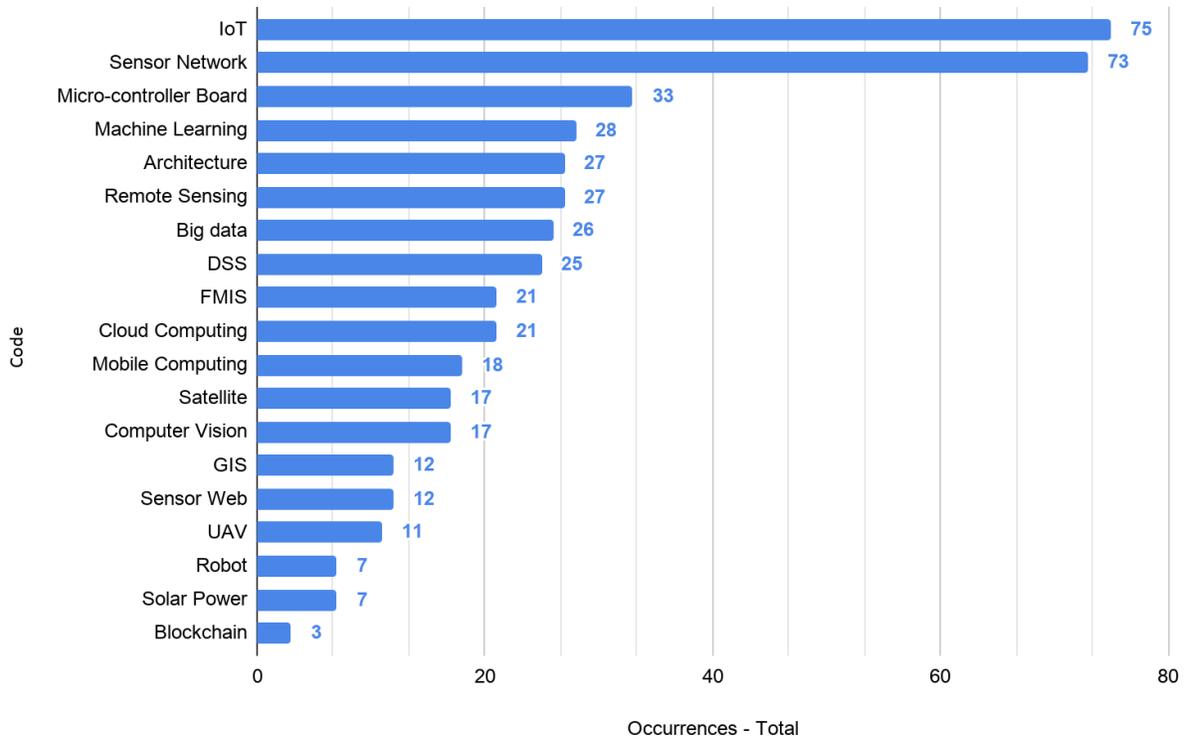

**Figure 6. Technology codes frequency**

Table 2 provides the technology codes and a systematic mapping to the extracted studies.

**Table 2. Technology Mapping**

| Code (Frequency) | Concept | Related Studies |
|---|---|---|
| IoT (75) | Internet of things refers to networks of digital smart devices - may be used as sensor networks, remote control, or automation | [1], [2], [4], [5], [7], [8], [9], [14], [16], [17], [20], [21], [22], [26], [27], [30], [31], [32], [38], [39], [40], [41], [42], [43], [44], [46], [47], [49], [50], [52], [53], [56], [58], [60], [62], [64], [65], [67], [68], [71], [72], [73], [74], [75], [76], [77], [79], [82], [88], [90], [92], [94], [95], [96], [103], [104], [106], [107], [111], [114], [115], [116], [121], [122], [126], [128], [129], [131], [132], [134], [135], [137], [140], [141], [142] |
| Sensor Network (73) | Hardware network for monitoring environmental conditions | [1], [2], [3], [4], [5], [6], [7], [8], [9], [11], [14], [16], [17], [20], [22], [25], [26], [27], [28], [30], [32], [36], [38], [39], [40], [42], [43], [44], [49], [50], [52], [53], [55], [56], [57], [62], [67], [68], [71], [72], [73], [75], [76], [78], [79], [80], [82], [85], [88], [89], [90], [91], [92], [94], [95], [96], [103], [104], [106], [107], [114], [115], [116], [121], [122], [123], [128], [129], [132], [135], [136], [137], [140], [141], [142] |
| Micro-controller Board (33) | Computer hardware used in digital agriculture technology - for example, Arduino and Raspberry Pi | [1], [5], [8], [9], [16], [22], [26], [27], [30], [32], [42], [43], [44], [57], [60], [65], [67], [73], [75], [76], [80], [82], [83], [84], [88], [92], [94], [103], [106], [124], [131], [135], [136] |
| Machine Learning (28) | Artificial intelligence and predictive modeling - Prediction and outcomes with data models, statistical learning | [3], [4], [6], [13], [31], [35], [36], [39], [43], [47], [48], [51], [54], [61], [68], [70], [84], [87], [90], [100], [101], [102], [104], [111], [120], [121], [131], [133] |
| Architecture (27) | Systems architecture or models | [14], [18], [20], [22], [25], [34], [37], [46], [49], [52], [53], [59], [66], [71], [85], [86], [89], [114], [115], [118], [128], [130], [132], [135], [139], [140], [141] |
| Remote Sensing (27) | Technology platforms that provide imaging and GPS - may include drones and satellites | [13], [15], [18], [21], [24], [29], [31], [41], [47], [49], [55], [63], [71], [74], [81], [82], [84], [88], [100], [104], [107], [119], [124], [126], [132], [138], [139] |
| Big Data (26) | Systems and processes associated with large amounts of data - includes data mining, data analysis, data processing | [10], [15], [21], [34], [36], [51], [56], [59], [61], [64], [80], [85], [93], [96], [97], [98], [99], [109], [113], [115], [118], [121], [125], [127], [130], [131] |
| DSS (25) | Decision Support System - A system connected to a database that provides information to help inform agricultural decisions | [6], [10], [15], [21], [22], [28], [43], [45], [47], [54], [63], [66], [69], [71], [74], [75], [80], [86], [93], [96], [118], [125], [127], [134], [135] |
| FMIS (21) | Farm Management Information Systems - MIS system used for the management/control of digital farm technologies | [17], [25], [37], [38], [46], [51], [55], [66], [69], [80], [85], [87], [89], [96], [97], [98], [99], [113], [128], [131], [132] |
| Cloud Computing (21) | Data and application resources available remotely through internet connections | [22], [29], [42], [43], [60], [64], [65], [71], [75], [77], [86], [89], [94], [99], [116], [123], [131], [134], [138], [139], [142] |
| Mobile Computing (18) | Hardware and software systems for mobile smart devices | [3], [11], [19], [20], [22], [28], [34], [45], [50], [54], [72], [78], [79], [81], [85], [118], [132], [140] |
| Satellite (17) | Orbital technology platforms used for GPS, communication, and imaging | [9], [15], [21], [29], [41], [56], [61], [63], [71], [74], [78], [81], [90], [104], [119], [126], [132] |
| Computer Vision (17) | Training a computer system to interpret and understand the visual world | [3], [6], [13], [47], [48], [54], [61], [68], [70], [84], [90], [100], [101], [102], [104], [111], [133] |
| GIS (12) | Geoinformatics, geospatial, and geography Information Systems | [25], [29], [33], [34], [63], [80], [98], [108], [120], [125], [126], [132] |
| UAV (11) | Unmanned Aerial Vehicle - includes remote controlled and automated drones or aircraft | [13], [18], [19], [21], [24], [31], [47], [49], [82], [100], [119] |
| Sensor Web (11) | The software components or architecture for sensor networks often used in environmental monitoring | [25], [49], [53], [60], [77], [107], [123], [134], [138], [139], [142] |
| Robot (7) | Tractor or mechanical farm device - may include field robots, self-driving tractors, and enhanced farm machinery | [19], [48], [55], [83], [84], [112], [124] |
| Solar Power (7) | Technology used in the collection, conversion, and storage of solar energy | [26], [42], [52], [76], [108], [122], [135] |
| Blockchain (3) | Software and systems for decentralized distributed digital ledger used to record transactions across multiple computers | [11],[46],[105] |

To understand agriculture technology in an open context, we explored the co-occurrence of open context coding with technology coding. The text below explores the top three technology codes that co-occur with the open context (Refer to Figure 7 for network diagram).

**Open Software** appeared in studies focused on IoT (55), studies with Sensor Network (53), and studies with Remote Sensing (22). IoT often provides software infrastructure and analysis tools for networked systems of local sensors and remote data collection technologies. The high frequencies of co-occurrence for these technologies with open software implies that open software is common in networked systems and is predominantly used in systems that measure environmental conditions.

**Open Data** appears along with DSS (21), Micro-controller Board (21), and Sensor Network (18). These studies explored how data collected in local sensor networks could be shared. The focus was often on the physical collection and sharing of data or supplementing local information with freely available open data. Farmers use open data to inform decisions and share data on soil quality, water content, disease identification for crops, and improve accuracy measures related to crop yield optimization.

**Open System** often appeared with IoT (30), Sensor Network (30), and Micro-controller Board (16). Studies with combinations of these codes describe large systems or interconnected systems of open software and open hardware that measure or control environmental factors.

**Open Hardware** appeared with IoT (33), Sensor Network (31), and Micro-controller Board (24). Studies with combinations of these codes describe the physical infrastructures of IoT sensor networks. Collections of open hardware include physical sensors and Arduino or Raspberry Pi[13] boards.

**Open Standard** appeared with IoT (13), Sensor Network (13), and Big Data (9). Studies with combinations of these codes focused on interdependencies and sharing data between systems. Specifically, these papers focus on the standardization of shared data and the shared rules for interdependent systems.

**Open Design** appeared with Sensor Network (6), IoT (3), and Architecture (3). Studies with open design concepts often presented a more holistic approach to building systems. Collaborative and transparent design processes were a focal concern. This data shows that open design research is relatively immature in open agriculture research.

---

13 https://www.raspberrypi.org/

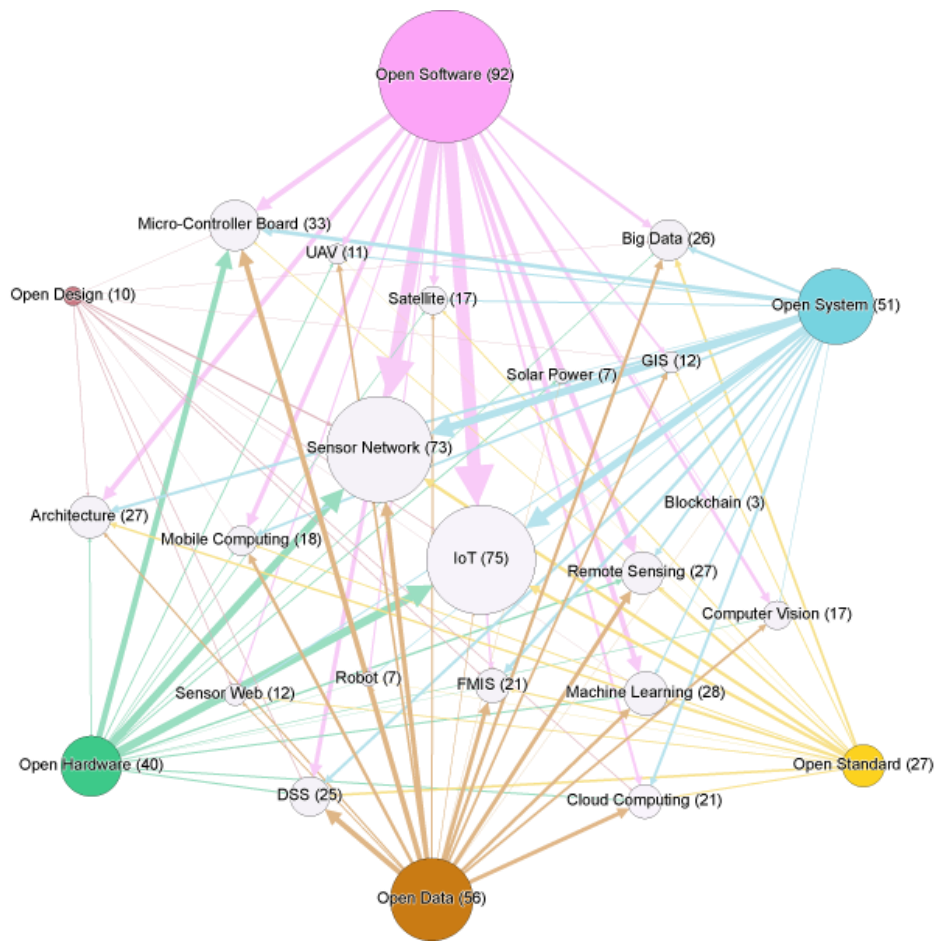

**Figure 7. Open Context and Technology co-occurrence network diagram**

The number attached to each code in the diagram represents the number of its unique occurrences. The size of each arrow in the graph represents the number of co-occurrences between codes.

### 3.4.3 Research Question 3: Open Applications

*What are the different applications of open source agriculture technology?*

We found 19 open agriculture digital technology applications that were categorized as either **Knowledge Management** or **Production Management** applications. The knowledge management applications were primarily concerned with creating and analyzing data. The production management applications were primarily concerned with controlling or automating processes based on the knowledge management applications however few studies included the creation of both these applications.

**Knowledge Management** applications concerned measuring, managing, and presenting data that could help inform agriculture decisions. Many of these codes were about the measurement of environmental conditions such as moisture levels (Hydrological (41)), weather (Meteorological (24)), and plant growth (Vegetation (24)).

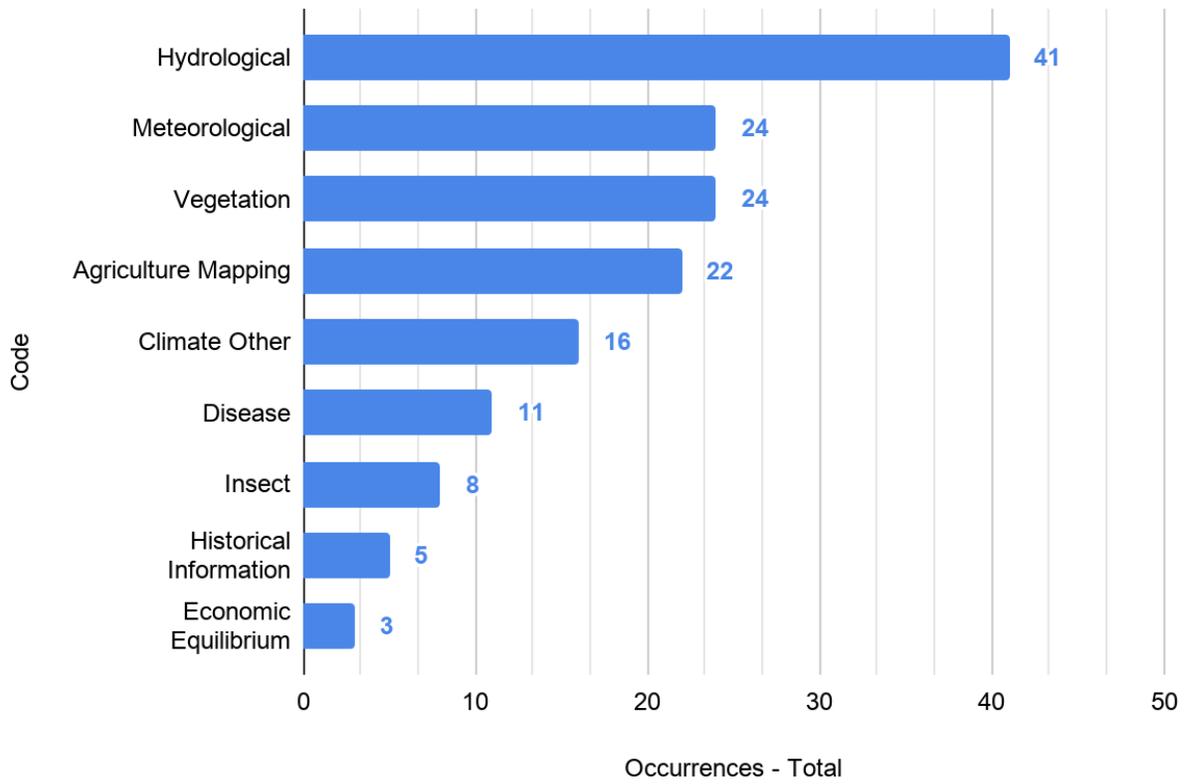

**Figure 8. Application - Knowledge Management coding frequency**

Table 3 provides the knowledge management application codes and a systematic mapping to the extracted studies.

**Table 3. Application - Knowledge Management mapping**

| Code (Frequency) | Concept | Related Studies |
|---|---|---|
| Hydrological (41) | Monitoring - precipitation levels, ground water level, soil moisture | [2], [5], [8], [9], [16], [25], [26], [27], [31], [32], [33], [35], [43], [44], [52], [57], [61], [62], [67], [68], [71], [72], [75], [76], [77], [80], [92], [95], [104], [106], [107], [108], [114], [115], [129], [134], [135], [136], [138], [139], [142] |
| Meteorological (24) | Weather (usually temperature and precipitation), air pressure, solar radiation | [5], [8], [9], [20], [21], [32], [35], [40], [43], [45], [62], [65], [67], [76], [80], [86], [95], [106], [107], [115], [116], [122], [129], [135] |
| Vegetation (24) | Concerned with measuring and identifying plant matter | [6], [13], [15], [18], [21], [31], [41], [47], [48], [59], [61], [63], [68], [70], [84], [90], [101], [102], [105], [111], [118], [119], [125], [133] |
| Agriculture Mapping (22) | Vegetation detection, calculation of standard vegetation indices, canopy greenness estimation, land surface temperature estimation and time-series analytics, typology, cartography | [15], [18], [25], [29], [31], [34], [41], [56], [61], [63], [69], [81], [96], [100], [104], [108], [109], [119], [125], [126], [132], [139] |
| Climate Other (16) | Climate that is not explicitly meteorological or hydrological for example - soil temp, Ph, Co2 | [2], [9], [16], [32], [53], [58], [62], [67], [74], [79], [80], [97], [104], [107], [114], [135] |
| Disease (11) | Diseases that negatively affect crops | [3], [6], [18], [36], [47], [63], [85], [89], [94], [107], [122] |
| Insect (8) | Prediction and measurement of pests affecting crops | [3], [20], [28], [36], [39], [47], [63], [91] |
| Historical Information (6) | Historic information about farming - includes historic crop prices, yields, past weather, disease, not real time monitoring | [45], [69], [81], [86], [87], [116] |
| Economic Equilibrium (3) | Economic factors - includes crop prices and supply and demand information | [33], [87], [94] |

To understand knowledge management applications technology in an open context, we explore the co-occurrence of open context coding with knowledge management application coding. The text below explores the top three knowledge management application codes that co-occur with the open context (Refer to Figure 9 for network diagram).

In our systematic mapping study, **Open Software** appears in studies coded with Hydrological (30), studies coded with Vegetation (17), and studies coded with Agriculture Mapping (16). These co-occurrence trends indicate that open software is used in knowledge management applications that connect environmental and crop data to geo-locations.

**Open Data** often appeared with Meteorological (14), Climate Other (12), and Agriculture Mapping (11). Similar to open software, analysis showed that open data is used in knowledge management applications that connect environmental data to geographic locations. These studies often specifically reference the Open Geospatial Consortium (OGC) [14] which provides open source geospatial data and standards frequently used in agricultural settings.

**Open System** was often coded with Hydrological (15), Agriculture Mapping (9), and Meteorological (8). Similar to open software and open data, these co-occurrence trends indicate that open agriculture systems include multiple applications related to mapping environmental data to geo-locations.

---

14  https://www.ogc.org/

**Open Hardware** appeared with Climate Other (9), Meteorological (12), and Hydrological (18). This co-occurrence trend is an indication that regarding knowledge management, open hardware devices are primarily used to measure local environmental factors

**Open Standard** appeared with Hydrological (11), Meteorological (7), and Agriculture Mapping (7). The presence of similar co-occurrences helps explain that open systems and open standards are used together in knowledge management applications. Supporting information is provided here that the development of Open Standards influences the methods used to measure and map environmental factors.

**Open Design** had a very low frequency of occurrence and co-occurrence making it difficult to describe the co-occurrences as trends.

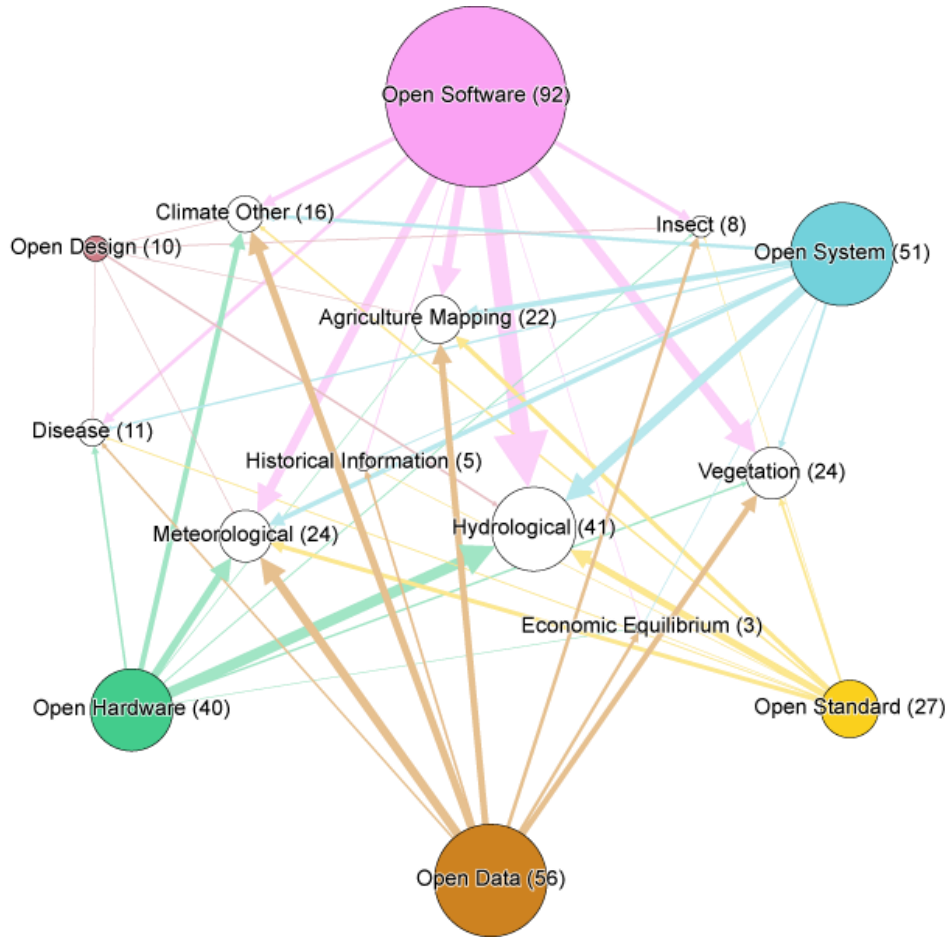

**Figure 9. Open Context and Application - Knowledge Management co-occurrence network diagram**

The number attached to each code in the diagram represents the number of its unique occurrences. The size of each arrow in the graph represents the number of co-occurrences between codes.

**Production Management** applications were concerned with manipulating technology and environmental factors. While knowledge management was about measuring the environment, production management was often about controlling a situation. Most of these codes were about control over agricultural systems (Management & Control (31)), moisture levels (Irrigation (30)), or manipulation of conditions to maximize crop yield (Crop Yield Optimization (28)).

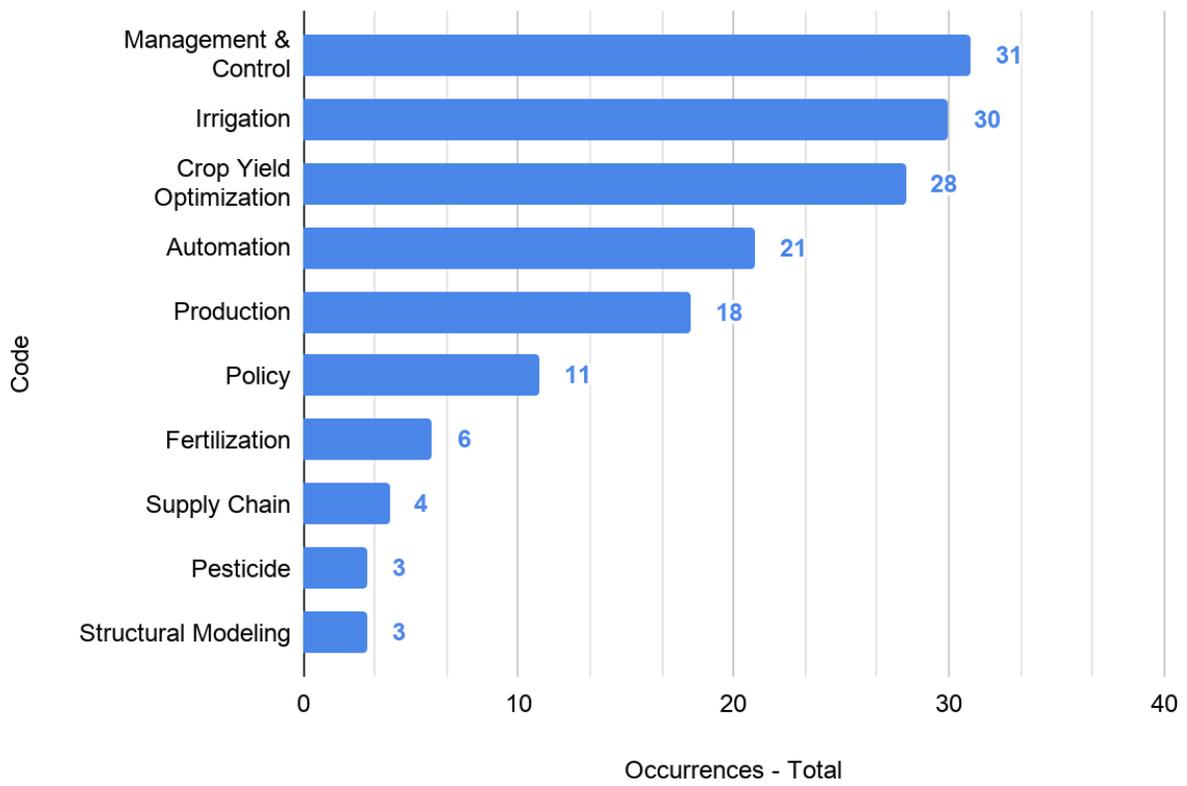

**Figure 10. Application - Production Management coding frequency**

Table 4 provides the production management application codes and a systematic mapping to the extracted studies.

Table 4. Application - **Production** Management mapping

| Code (Frequency) | Concept | Related Studies |
|---|---|---|
| Management & Control (31) | Software, hardware, or systems used for the Management and control of agriculture technology | [1], [4], [20], [21], [24], [31], [32], [33], [39], [43], [47], [48], [57], [66], [67], [68], [88], [89], [105], [106], [108], [114], [115], [116], [118], [124], [132], [134], [135], [136], [142] |
| Irrigation (30) | Method to increase crop yield and prediction - land use and climate smart applications | [1], [5], [9], [20], [21], [26], [31], [32], [33], [43], [44], [57], [67], [68], [71], [75], [77], [88], [92], [94], [103], [106], [108], [114], [115], [116], [132], [134], [136], [142] |
| Crop Yield Optimization (28) | Control of factors to increase production - for example irrigation, fertilization, climate | [13], [14], [17], [18], [23], [28], [33], [35], [39], [41], [42], [45], [49], [53], [54], [63], [64], [66], [71], [78], [89], [98], [99], [120], [128], [129], [133], [137] |
| Automation (22) | Running software and hardware processes automatically | [7], [18], [19], [24], [30], [32], [43], [49], [55], [67], [83], [84], [89], [94], [106], [112], [116], [121], [131], [136], [137], [140] |
| Production (17) | System is focused on the production of a resource | [4], [33], [37], [46], [51], [55], [56], [74], [78], [91], [93], [95], [117], [122], [129], [136], [137] |
| Policy (11) | Governance, legal, and defined coordination mechanisms | [10], [12], [22], [23], [28], [78], [86], [93], [109], [110], [130] |
| Fertilization (6) | Application of fertilizer - boosting the growth and output of plants | [21], [24], [47], [67], [83], [132] |
| Supply Chain (5) | System is focused on distribution of a resource | [11], [17], [46], [86], [131] |
| Pesticide (3) | Management and application of chemicals to eliminate or mitigate pests affecting crops | [20], [39], [47] |
| Structural Modeling (3) | Plans and designs for physical objects - includes 3D printing | [91], [117], [122] |

To understand production management applications technology in an open context, we explore the co-occurrence of open context coding with production management application coding. The text below explores the top three production management application codes that co-occur with the open context (Refer to Figure 11 for network diagram).

In our systematic mapping study **Open Software** appears in studies coded with Management & Control (26), studies coded with Irrigation (24), and studies coded with Crop Yield Optimization (20). These co-occurrence trends indicate that Open Software is utilized in production management applications that manage environmental conditions to improve crop yield.

**Open Data** was often coded with Irrigation (12), Crop Yield Optimization (11), and Automation (7). Many of these co-occurrences resulted from the use of open data to improve existing production practices. For example, open data may be used to automate applications such as irrigation to improve crop yield.

**Open System** often appeared with Irrigation (13), Management & Control (11), and Crop Yield Optimization (8). Similar to open software, studies coded with open systems were concerned with managing environmental conditions to maximum crop yield.

**Open Hardware** often appeared with Irrigation (15), Management & Control (14), and Automation (10). These co-occurrence trends highlight agricultural technology that controls and automates irrigation systems, a primary concern in agriculture.

**Open Standard** often appeared with Management & Control (7), Irrigation (6), and Crop Yield Optimization (6). The presence of similar co-occurrences describes that open systems and open standards are used together in production management applications. Supporting information is provided here that the development of open standards influences the management and control of environmental factors to improve crop yield.

**Open Design** appeared at a very low frequency of occurrence and co-occurrence making it difficult to describe the co-occurrences as trends.

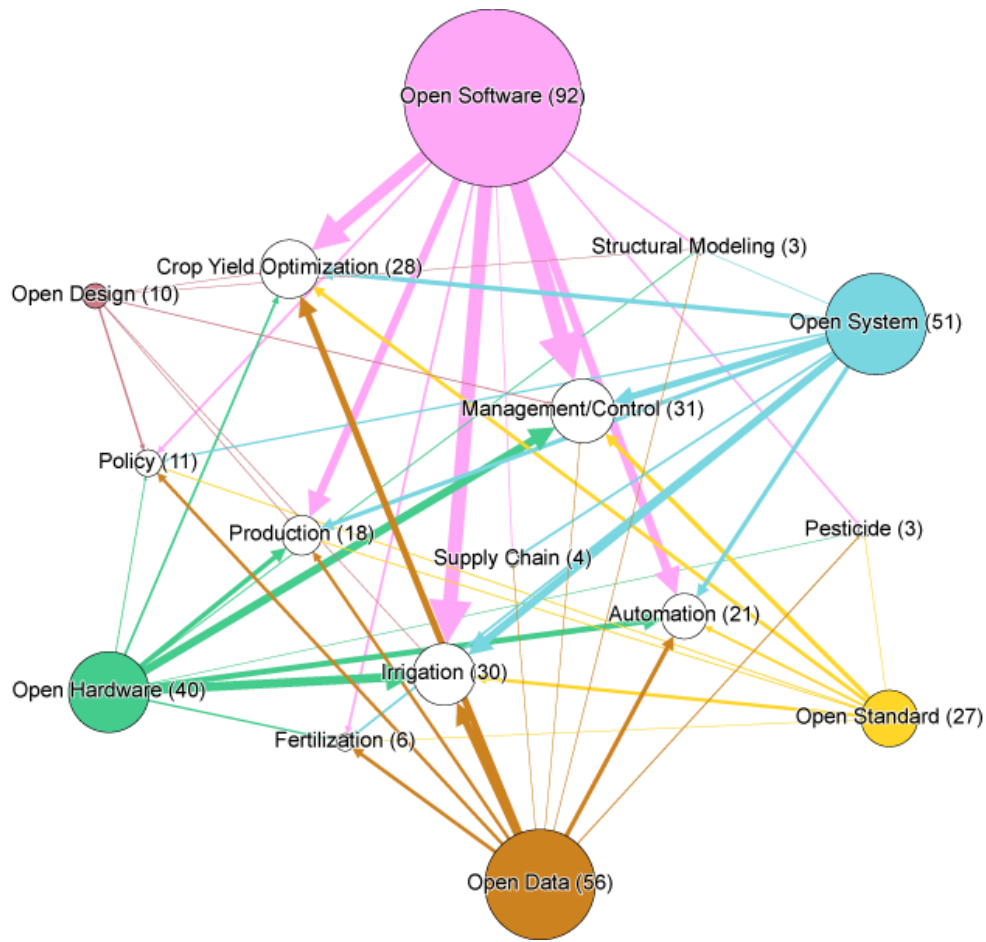

**Figure 11. Open Context and Application - Production Management co-occurrence network diagram.**

The number attached to each code in the diagram represents the number of its unique occurrences. The size of each arrow in the graph represents the number of co-occurrences between codes.

### 3.4.4    Research Question 4: Agricultural Context

*What are the different **agricultural contexts**?*

While all the studies that we analyzed had an agricultural context, several had a very specific agricultural context or focal area. We observed nine specialized agriculture contexts (refer to Figure 12 and Table 5).  In this section, we describe the frequency of these specialized agricultural contexts and explore what types of applications were present in the studies.

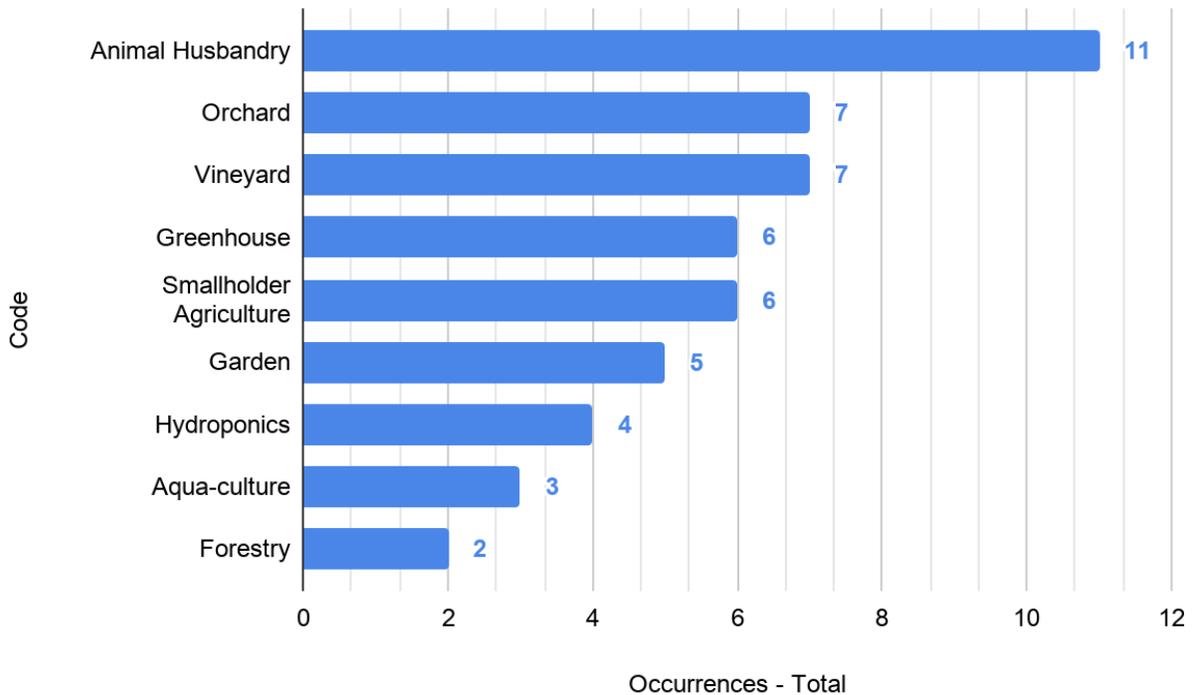

**Figure 12. Agriculture Context coding frequency**

**Animal Husbandry** was coded in 11 studies as the agricultural context for digital agriculture technology research. This included systems that monitored the activity of livestock, analyze their behavior, track their health, and identify animal species. Technology that is used in this regard includes computer vision, wearable sensors, cameras, accelerometer, smartphones, and shareable open data.

**Orchards** was coded in 7 studies. Research centered on field robots, decision support systems, and IoT. Most of the papers involved orchards that grow fruit, including peaches and oranges. One proof-of-concept paper involved an orchard that was completely simulated to accelerate robotics research.

**Vineyard** was a subject of 7 studies, which centered around measurement of the environment and detection of hazards. Some papers included the use of an Unmanned Aerial Vehicle (UAV) to measure a vineyard while other papers focused on measuring data points like water and disease.

**Greenhouses** provided a research setting for 6 studies. Much of this research focused on measuring water content or controlling irrigation. An interesting topic introduced with greenhouses was the security aspect of greenhouse sensors.

**Smallholder Agriculture** was coded in 6 studies. These studies mostly focused on the future of agriculture and the inclusion of precision agriculture in small or developing nations.

**Garden** was coded in 5 studies. Much of the focus in this space was on self-sufficiency and urban farming.

**Hydroponics** was coded in 4 studies. These papers generally involved sensors and irrigation systems. Some papers containing hydroponics focused on printable gardens and fabrication.

**Aqua-culture** was coded in 3 studies. Topics included mapping pools, monitoring inventory (shrimp), and the use of robots to clean the pools.

**Forestry** was coded in 2 related studies. One explored eCommerce for forestry while the other described climate-smart land use.

**Table 5. Agriculture Context Mapping**

| Code (Frequency) | Concept | Related Studies |
|---|---|---|
| Animal Husbandry (11) | The science of breeding and caring for farm animals | [2], [7], [10], [22], [46], [58], [73], [85], [91], [121], [141] |
| Orchard (7) | Tree farms that produce fruit or nuts | [55], [69], [75], [82], [112], [118], [129] |
| Vineyard (7) | Farm used to produce grapes or other vine-related plants | [34], [69], [76], [82], [119], [122], [123] |
| Greenhouse (6) | Structurally confined farming | [50], [51], [53], [79], [134], [140] |
| Smallholder Agriculture (6) | Farms with limited resources compared to other farmers - for example, subsistence farming and small crop yield | [23], [28], [63], [108], [136], [137] |
| Garden (5) | Urban gardens, connected gardens, and potted gardens | [57], [68], [88], [95], [117] |
| Hydroponics (4) | Growing plants without soil in a controlled environment - Controlled fertilization, hydrological, meteorological | [4], [92], [117], [134] |
| Aquaculture (3) | Fisheries and water-based farms | [2], [126], [127] |
| Forestry (2) | The science or practice of planting, managing, and caring for forests | [20], [23] |

# 4   Discussion

This section presents our insight on the results of the systematic mapping study and implications for future work. We found six different open source concepts in the studies we analyzed. Open source has become quite common in many technological fields thus it is not surprising to find research in agriculture digital technology that references and utilizes open source technologies. Indeed, some technologies are primarily associated with open source such that the dominant or most utilized versions of the technology are not proprietary. Blockchain, for example, is primarily an open software and open standards technology. In our analysis we only coded open source that was explicitly referenced. What we found is that most of the studies utilized open source technologies but were not necessarily building or contributing back to open source projects. This is highlighted in the low frequencies of open design coding we observed. Viewing open source as primarily a licensing designation allows researchers to utilize free software as a component of their design stream but keeps it separate from the design process. This separation means that there is likely a disconnect between design theory and design outcomes (Lukyanenko & Parsons, 2020) because component open source technologies may have different design goals and theoretical underpinnings. This disconnect is troubling for validity of open agriculture research but offers an opportunity for Information Systems researchers.  Future open agriculture research needs to move past consumption of open technologies and focus on open design concepts.

Mirroring other disciplines, open hardware technologies are not as mature as open software and data systems in open agriculture research. We primarily observed open hardware in the use of micro-controller boards for the automation and control of environmental factors. However, our conceptualization of open hardware also included robots, drones, and other digitally enhanced physical technologies which we observed in only a few studies. Agriculture equipment is quite expensive and must often be maintained in near continuous uptimes. Open hardware technologies such as open source tractors and robots have received positive attention from farmers and media (Hibbets, 2011) but based on our systematic mapping, academic research has yet to fully embrace the topic. With rising costs associated with the purchase and repair of agriculture equipment, open hardware technologies have the potential to disrupt proprietary agriculture digital technology and have implications for the right-to-repair movement, which is trying to reduce farmer dependence on corporately controlled technology. Future open agriculture research should explore the open design of hardware technologies including tractors, drones, and robots.

Many of the knowledge management applications we observed were concerned with measuring local data points and combining the data with external data. Similarly, many of the production management applications were concerned with controlling local processes based on external data. The main exception to this observation was the specialized agricultural contexts of greenhouses and hydroponics. In these studies, the line between knowledge management applications and production management applications was often blurred. We view this distinction as a single purpose focus versus an ecosystem focus. Open systems can include many different technologies. While much of the research we reviewed included complementary and interdependent software and hardware, the application focus was often localized into a single purpose (i.e., measurement of an environmental variable to inform decision making or control of an environmental variable). Along with many technologies open systems can incorporate many applications including knowledge management and production management. Future research should explore open agriculture more holistically, accounting for complementary and interdependent software ecosystems, and exploring how farmers use and interact with them. Of specific interest to Information Systems researchers in open agriculture research are farm management Information Systems (FMIS) and Enterprise Resource Planning (ERP) systems.

# 5   Limitations and Threats to Validity

There are limitations of this systematic mapping study. First, systematic reviews are prone to bias, which is shown in the inclusion or exclusion of studies (Higgins et al., 2019). We attempted to mitigate this issue by performing a greedy search (2,421 study results) and having two researchers manually verify the relevance of each study which resulted in a considerably smaller number of results (241 relevant studies). Similarly, this study is limited by the bias of the searched databases. The databases we selected represent a broad spectrum of journals and studies, but they do not encompass all the possible venues for open agriculture technology research. To minimize this bias, we added a Google Scholar search to our greedy search to be more inclusive and not miss any key studies. Second, the search outcomes are based on the specific search logic we developed and used to find the relevant studies. There might be other terms that might be used to represent open source and are not captured in this research. Thus, this research is not a complete representation of open source work in the agriculture domain. Third, this work provides a systematic mapping and not a quality assessment of the studies. Future work can explore the quality of the studies included in this systematic mapping. Fourth, in this work, we have coded the studies into different themes. Coding by different authors introduces bias based on subjective judgment. We attempted to mitigate this bias through parallel coding of the studies by two researchers. In case of any discrepancies, the third researcher served as a tiebreaker.

# 6   Conclusion

Open Agriculture is an emergent field with substantial opportunity for Information Systems researchers. This systematic mapping study was conducted to develop an understanding of the open agriculture digital technology. In this mapping study, we performed a rigorous review of 2421 papers, which resulted in 142 final selections. The papers were then categorized with an open coding technique in which relevant themes in the studies were identified. We have provided a systematic mapping of open agriculture research along with the results of our open coding and our insights. Agriculture digital technology is crucial to the global community and the convergence between agriculture digital technology and open technologies is growing. While open agriculture digital technology is being consumed by farmers and researchers, the design implications have yet to be explored, providing a research opportunity for Information Systems researchers.

# Appendix A: Selected Studies